\documentclass[preprints,article,accept,moreauthors, pdflatex]{Definitions/mdpi} 
\firstpage{1} 
\pubvolume{xx}
\issuenum{1}
\articlenumber{5}
\pubyear{2019}
\copyrightyear{2019}
\history{Received: date; Accepted: date; Published: date}

\pdfoutput=1

\usepackage{bbding}
\usepackage{pifont}
\usepackage{wasysym}
\usepackage{amssymb}

\Title{Deep Encoder-decoder Adversarial Reconstruction (DEAR) Network for 3D CT from Few-view Data}

\Author{Huidong Xie $^{1}$, Hongming Shan $^{1}$  and Ge Wang $^{1}$*}

\AuthorNames{Huidong Xie, Hongming Shan and Ge Wang}

\address{%
$^{1}$ \quad Biomedical Imaging Center, Department of Biomedical Engineering, Center for Biotechnology \& Interdisciplinary Studies, Rensselaer Polytechnic Institute, 110 Eighth Street, Troy NY 12180, USA}

\corres{Correspondence: wangg6@rpi.edu}

\abstract{X-ray computed tomography (CT) is widely used in clinical practice. The involved ionizing X-ray radiation, however, could increase cancer risk. Hence, the reduction of the radiation dose has been an important topic in recent years. Few-view CT image reconstruction is one of the main ways to minimize radiation dose and potentially allow a stationary CT architecture. In this paper, we propose a deep encoder-decoder adversarial reconstruction (DEAR) network for 3D CT image reconstruction from few-view data. Since the artifacts caused by few-view reconstruction appear in 3D instead of 2D geometry, a 3D deep network has a great potential for improving the image quality in a data-driven fashion. More specifically, our proposed DEAR-3D network aims at reconstructing 3D volume directly from clinical 3D spiral cone-beam image data. DEAR is validated on a publicly available abdominal CT dataset prepared and authorized by Mayo Clinic.  Compared with other 2D deep-learning methods, the proposed DEAR-3D network can utilize 3D information to produce promising reconstruction results. }


\begin{document}


\section{Introduction}
X computed tomography (CT) is one of the most essential imaging modalities widely used in clinical practices \cite{brenner_computed_2007}. Even though CT brings overwhelming healthcare benefits to patients, it could potentially increase the patients' cancer risk due to the involved ionizing radiation. The data from the National Lung Screening Trial indicate that annual lung cancer screening with low-dose CT could significantly reduce lung cancer-related mortality \cite{shan_3-d_2018-1}. If the effective dose of a routine CT examination is reduced to less than 1 mSv, the long-term risk of CT scanning can be negligible. In the past years, numerous deep-learning-based CT denoising methods were proposed to reduce radiation dose with excellent results \cite{chen_low-dose_2017-1, shan_competitive_2019-1, chen_low-dose_2017}. In parallel, few-view CT is also being actively investigated to reduce the radiation dose, especially for breast CT \cite{glick_breast_2007-1} and C-arm CT \cite{wallace_three-dimensional_2008}.

Few-view CT is a challenging problem. Due to the requirement imposed by the Nyquist sampling theorem \cite{jerri_shannon_1977}, reconstructing high-quality CT images from under-sampled projection data was previously considered an unsolvable problem. With sufficient projection data, analytical methods such as filtered back-projection (FBP) \cite{wang_approximate_2007} can be used for accurate image reconstruction. However, FBP will introduce severe streak artifacts when projection data are limited. Numerous iterative reconstruction algorithms were proposed to incorporate prior knowledge for suppressing image artifacts in few-view scans. Well-know methods include algebraic reconstruction technique (ART) \cite{gordon_algebraic_1970}, simultaneous algebraic reconstruction technique (SART) \cite{andersen_simultaneous_1984}, expectation maximization (EM) \cite{dempster_maximum_1977}, etc. Even though these iterative methods do improve image quality, they are usually time-consuming and still not able to produce clinically-acceptable results in many cases. Recently, with the assistance of graphics processing unit (GPU) and big data, deep learning has become a new frontier of tomographic imaging and gives new opportunities for few-view CT reconstruction \cite{wang_perspective_2016, wang_guest_2015}.

Deep learning has been now well recognized in the field of medical tomographic imaging \cite{greenspan_guest_2016}. Several methods were proposed to resolve few-view CT issues in a data-driven fashion. For example,  Jin et al. \cite{jin_deep_2017} proposed a U-net-based \cite{ronneberger_u-net:_2015} FBPConvNet to remove streak artifacts in the 2D image domain. Lee et al. used a similar U-net structure to eliminate artifacts in the sinogram domain \cite{lee_deep-neural-network-based_2019}. Chen et al. designed a LEARN network \cite{chen_learn:_2018} to map sparse sinogram data directly to a tomographic image, which combines the convolutional neural network \cite{lecun_convolutional_1995} and a classic iterative process under a data-driven regularization. Inspired by the FBP workflow, Li et al. published their iCT-NET \cite{li_learning_2019} to perform CT reconstruction in various special cases and consistently obtain decent results. Our recently published DNA network \cite{xie_dual_2019-1} addressed the few-view CT issue by learning a network-based reconstruction algorithm from sinogram data. But none of these proposed methods were designed to perform 3D image reconstruction, subject to potential loss in the 3D context.

In this paper, we propose a deep encoder-decoder adversarial reconstruction network (DEAR) for 3D CT from few-view data, featured by a direct mapping from a 3D input dataset to a 3D image volume. In diagnosis, radiologists need to extract 3D spatial information by looping adjacent slices and form contextual clues. Therefore, it is reasonable and even necessary to use 3D convolutional layers for maximally avoiding streak artifacts in a batch of adjacent reconstructed image slices. The main contributions of our DEAR-3D network are summarized as follows: 
\begin{enumerate}
\item[1)] DEAR-3D utilizes 3D convolutional layers to extract 3D information from multiple adjacent slices in a generative adversarial network (GAN) \cite{goodfellow_generative_2014}  framework. Different from reconstructing 2D images from 3D input data \cite{shan_3-d_2018-1}, DEAR-3D directly reconstructs a 3D volume, with faithful texture and image details; 
and 
\item[2)] An extensive comparative study was performed between DEAR-3D and various 2D counterparts to demonstrate the merits of the proposed 3D network.
\end{enumerate}

The rest of this paper is organized as follows. Sec.\ref{sec:method} introduces the DEAR-3D model, its 2D counterparts and the GAN framework utilized in the proposed model. Sec.\ref{sec:exp} describes our experimental design and results, in comparison with other state-of-the-art models for few-view CT. Finally, Sec.\ref{sec:conclusion} presents discussions and concludes this paper.

\section{Methodology}\label{sec:method}
3D CT image reconstruction can be expressed as follows:
\begin{equation}
I=R^{-1}(S)\label{eq1}
\end{equation}
where $I\in \mathbb{R}^{N_s\times N\times N}$ denotes a 3D image volume reconstructed from sufficient projection data, where $N$ and $N_s$ denote the width/height of input images and number of images acquired from a particular patient respectively. $S\in \mathbb{R}^{N_v\times N_d \times N_r}$ denotes the corresponding interpolated 3D sinogram from a spiral cone-beam scan, $N_v$, $N_d$ and $N_r$ denote the number of views, the number of detectors per row, and the number of detector rows respectively, and $R^{-1}$ is the inverse operator to reconstruct the CT image volume, such as a typical cone-beam reconstruction formula or algorithm \cite{schaller_efficient_1998,grangeat_mathematical_1991,grangeat_evaluation_1991, katsevich_improved_2004} when sufficient projection data are obtained. However, when the number of data (linear equations) is not sufficient to resolve all the unknown voxles in the few-view CT setting, streak artifacts will be introduced in the reconstructed images, and how to reconstruct high-quality images becomes highly non-trivial. Deep learning (DL) promises to extract features in reference to extensive knowledge hidden in big data. With a large amount of training data, task-specific and robust prior knowledge can be taken advantage of in establishing a relationship between few-view data/images and the corresponding full-view images. Such a deep network can be formulated in Eq. \eqref{eq2}, 
\begin{equation}
I_{\mathrm{FV}}=T(I_{\mathrm{SV}})\label{eq2}
\end{equation}
where $I_{\mathrm{FV}}$ and $I_{\mathrm{SV}}$ denote a 3D image volume reconstructed from sufficient projection data and the counterpart from insufficient few-view/sparse-view data, respectively, and $T$ denotes our DEAR-3D network to remove artifacts caused by few-view problem.

\subsection{Proposed Framework}
\begin{figure*}
\centering
\includegraphics[width=\textwidth]{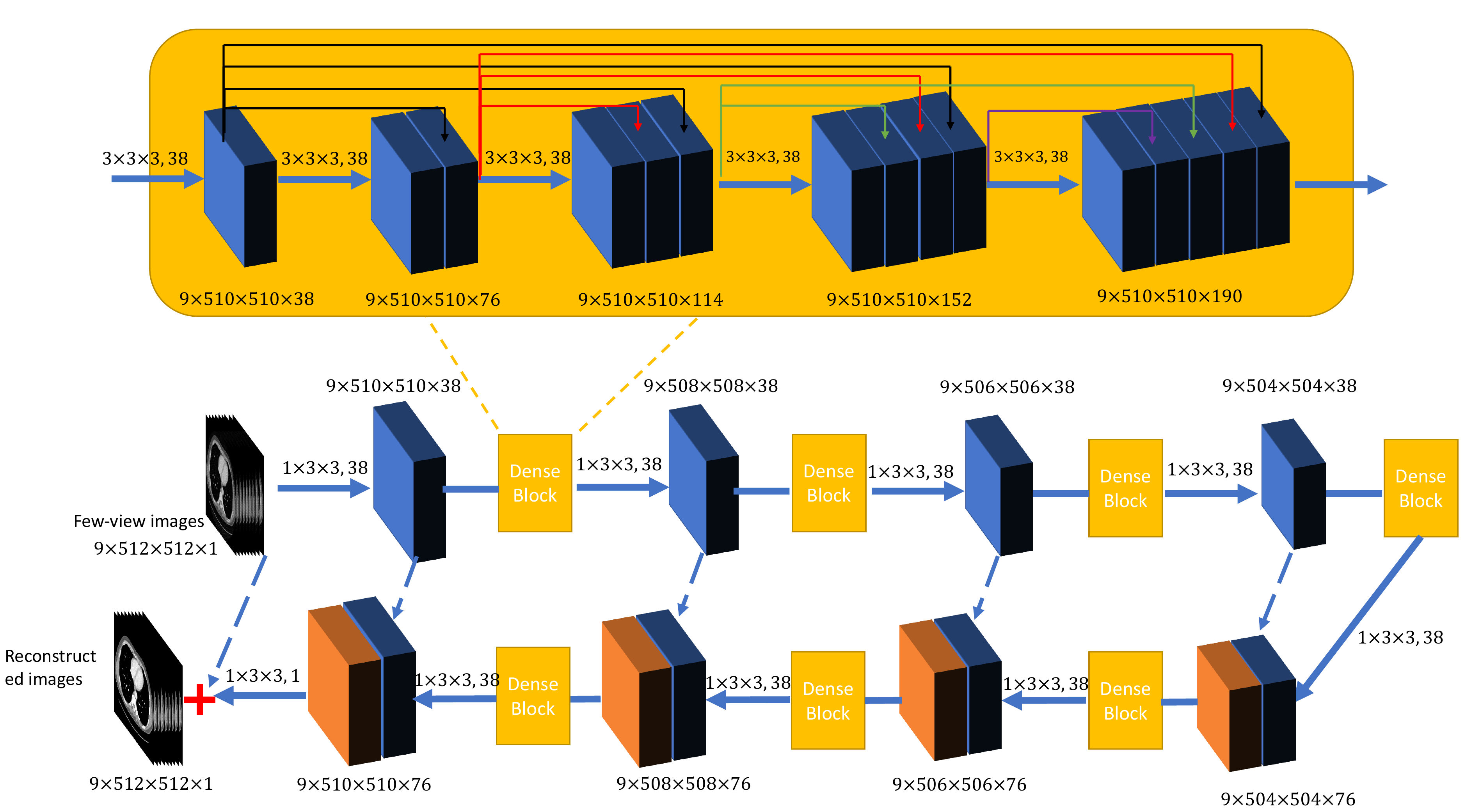}
\caption{The overall structure of the proposed generator network $G$ in the DEAR-3D network. Note that the numbers beside each rectangular block represent the dimensionality of the feature-maps, the dotted blue lines represent conveying paths. The images are the input and output examples for the proposed method. The numbers above each solid blue arrows are for the convolutional kernel and the number of filters in the format of $\mathsf{kernel\ size}\times \mathsf{number\ of\ filters}$. Solid black, red, green and purple arrows represent conveying paths inside dense blocks.}
\label{fig:Network_structure}
\end{figure*}

The overall network architecture is shown in Fig. \ref{fig:Network_structure}. The proposed DEAR-3D network is optimized in a Wasserstein Generative Adversarial Network (WGAN) framework\cite{arjovsky_wasserstein_2017}, which is currently one of the most advanced frameworks. In this study, the proposed framework consists of two components: a generator network $G$ and a discriminator network $D$. $G$ aims at directly reconstructing a 3D image volume from a batch of 3D few-view image slices. $D$ receives images from both $G$ and the ground-truth dataset, trying to distinguish whether the input is real. Both networks optimize themselves during the training process. If an optimized network $D$ can hardly distinguish fake images (from $G$) from real images (from the ground-truth dataset), then the generator network $G$ fools the discriminator $D$ successfully. By the design, the introduction of $D$ also helps to improve the texture of reconstructed images.

Different from the vanilla generative adversarial network (GAN) \cite{goodfellow_generative_2014}, WGAN replaces the logarithm term in the loss function with the Wasserstein distance, improving the training stability. In WGAN, the 1-Lipschitz function is assumed with weight clipping. However, Ref \cite{gulrajani_improved_2017-1} pointed out that weight clipping may be problematic in WGAN, and suggested to replace it with a gradient penalty term, which is used in our proposed method. Hence, the objective function of the GAN framework is expressed as follows:
\begin{equation}
\begin{aligned}
\min_{\theta_G}\max_{\theta_D}\underbrace{\{\mathbb{E}_{I_{\mathrm{SV}}}[D(G(I_{\mathrm{SV}}))]-\mathbb{E}_{I_{\mathrm{FV}}}[D(I_{\mathrm{FV}})]}_{Wasserstein\ distance}
+\lambda\underbrace{\mathbb{E}_{\bar{I}}[(\|\nabla(\bar{I})\|_2-1)^2]\}}_{gradient\ penalty} \label{eq3}
\end{aligned}
\end{equation}
where $I_{\mathrm{SV}}$ and $I_{\mathrm{FV}}$ represent few-view/sparse-view 3D image volume and full-view 3D image volume respectively, $\mathbb{E}_a[b]$ denotes the expectation of $b$ as a function of $a$, $\theta_G$ and $\theta_D$ denote the trainable parameters of the networks $G$ and $D$ respectively, and $\bar{I}=\alpha\cdot I_{\mathrm{FV}}+(1-\alpha)\cdot G(I_{\mathrm{SV}})$. $\alpha$ is uniformly sampled from the interval [0,1]. In other words, $\bar{I}$ represents another batch of 3D image slices between fake and real images. Furthermore, $\nabla(\bar{I})$ denotes the gradient of $\bar{I}$ with respect to $\theta_D$. Lastly, $\lambda$ is a parameter used to balance the Wasserstein distance and the gradient penalty. The networks $G$ and $D$ are updated in an iterative manner as suggested by \cite{goodfellow_generative_2014,gulrajani_improved_2017-1, arjovsky_wasserstein_2017}.

\subsection{Generator Network}
The input to the generator $G$ is a batch of 3D image slices with dimensionality of $N_b \times N_s\times N \times N$ where $N_b$, $N_s$, and $N$ denote the batch size, number of adjacent input slices and dimension of each input image slice. Intuitively, $N_s$ should be equal to the total number of image slices of a particular patient, and tissues in all the different 2D planes should relate to each other. However, it is not practical due to an extremely large memory cost. Hence, $N_s$ is experimentally adjusted to 9. The structure of the generator  $G$ is inspired by the U-net \cite{ronneberger_u-net:_2015}, originally proposed for biological image segmentation. Since then, the U-net has been utilized for various applications in the field of medical imaging. For example, \cite{chen_low-dose_2017-1, shan_3-d_2018-1} used U-net with conveying paths for CT image denoising, \cite{xie_dual_2019-1, jin_deep_2017, lee_deep-neural-network-based_2019} applied U-net for few-view CT, and \cite{donoho_compressed_2006} for compressed sensing MRI . In DEAR, $G$ is a revised U-net with conveying paths and built in reference to DenseNet \cite{huang_densely_2016}. The generator  $G$ consists of 4 3D convolutional layers for down-sampling and 4 3D transpose convolutional layers for up-sampling a batch of 3D image slices. The dimension of 3D kernel for down-sampling and up-sampling is set as $1\times 3\times 3$. In the original U-net \cite{ronneberger_u-net:_2015}, stride of 2 was used in each down-sampling or up-sampling layer to extract features in different dimensions for segmentation. However, for image reconstruction, down-sampling input images severely may result in a compromised performance because convolutional layers may not be able to recover the images from low-dimensional feature maps accurately. Therefore, stride of 1 is used in all the convolutional layers of $G$ and zero-padding is not applied in down-sampling and up-sampling layers. A rectified linear unit (ReLU) activation function is used after each 3D convolutional layer.

A dense block is added after each down-sampling and up-sampling layer. Each dense block contains 5 3D convolutional layers to extract 3D image features from the input feature maps. Note that zero-padding is used in all 3D convolutional layers to maintain the dimensionality of the input feature maps. Inspired by ResNet \cite{he_deep_2015}, shortcuts are applied to connect early and current feature maps, allowing gradients to flow directly to the current layer from the corresponding earlier layer. Different from ResNet, DenseNet further improves the information flow between layers by connecting all the earlier feature maps to the current layer. Consequently, the $l^{th}$ layer receives all the feature maps from all previous layers, $x_0$, $x_1$, $x_2$, ... , $x_{l-1}$, as the input:
\begin{equation}
x_l = T_l([x_0, x_1, x_2, \ldots, x_{l-1}])\label{eq4}
\end{equation}
where $[x_0,x_1,x_2,\ldots,x_{l-1}]$ represents the concatenation of all the feature-maps produced by the layers $0, 1, 2,\ldots,l-1$, $T_l$ denotes the operation performed by the $l^{th}$ layer, and is defined as a composite function of a 3D convolutional operation and a ReLU activation. The kernel size and stride are set as $3\times 3\times 3$ and $1$ respectively for all the 3D convolutional layers in the proposed dense block. Note that the the purpose of DEAR-3D is to learn the inverse amplitude of artifacts in the input images, and therefore input images are directly added to the last convolutional layer as presented in Fig. \ref{fig:Network_structure}.

\subsection{Discriminator Network}
The discriminator network $D$ takes input from either $G$ or the ground-truth dataset, trying to classify whether the input images are real. In DEAR-3D, the discriminator network has 6 convolutional layers with 64, 64, 128, 128, 256, 256 filters and followed by 2 fully-connected layers with the numbers of neurons 1024 and 1 respectively. The leaky ReLU activation function is used after each layer with a slope of 0.2 in the negative part. 3D convolutional layers with $3\times3\times 3$ kernel dimension and zero-padding are used for all convolutional layers. Stride is set to 2 for all the layers.

\subsection{Objective Functions for Generator}

\begin{figure}
\centering
\includegraphics[width=\textwidth]{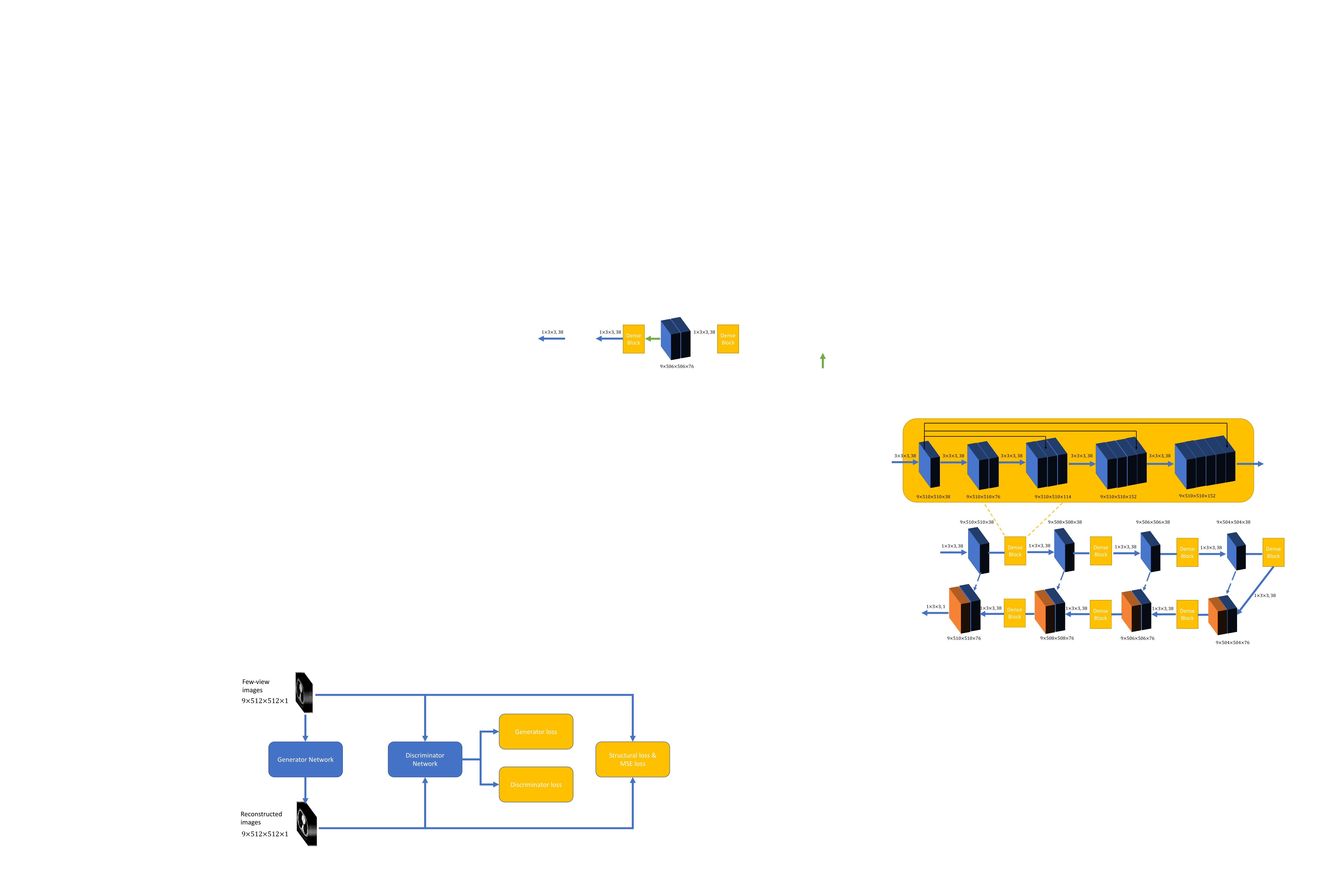}
\caption{Objective functions used to optimize the proposed DEAR-3D network.}
\label{fig:loss}
\end{figure}

This subsection introduces and evaluates different objective functions used for few-view CT artifact reduction. As shown in Fig. \ref{fig:loss}, a composite objective function is used to optimize DEAR-3D.

\subsubsection{MSE Loss}

The mean-squared-error (MSE) \cite{chen_low-dose_2017-1,wolterink_generative_2017,wang_mean_2009} is a popular choice for denoising and artifact-removal applications  \cite{wang_mean_2009}. Nevertheless, it could lead to over-smoothing images \cite{zhao_loss_2017-1}. Moreover, MSE is not sensitive to image texture and  assumes background noise is white Gaussian noise independent of local image features \cite{zhou_wang_image_2004}. The MSE used in the proposed method is expressed as follows:
\begin{equation}
L_2=\frac{1}{N_b\cdot N_s\cdot N \cdot N}\sum_{i=1}^{N_b}\|Y_i-X_i\|^2_2\label{eq5}
\end{equation}
where $N_b$ $N_s$ and $N$ denote the number of batches, number of input slices and image width/height respectively, $Y_i$ and $X_i$ represent the ground-truth 3D image volume and 3D image volume reconstructed by $G$ respectively.

\subsubsection{Structural Similarity Loss}

To overcome the disadvantages of MSE loss and acquire visually superior images, the structural similarity index (SSIM) \cite{zhou_wang_image_2004} is introduced in the objective function. SSIM measures structural similarity between two images. The convolution window used to measure SSIM is set to $11\times 11$. The SSIM is expressed as follows:
\begin{equation}
SSIM(Y,X)=\frac{(2\mu_Y\mu_X+C_1)(2\sigma_{YX}+C_2)}{(\mu_Y^2+\mu_X^2+C_1)(\sigma_Y^2+\sigma_X^2+C_2)}  \label{eqn:6}
\end{equation}
where $C_1=(K_1\cdot R)^2$ and $C_2=(K_2\cdot R)^2$ are constants used to stabilize the formula if the denominator is too small, $R$ stands for the dynamic range of voxels values, $K_1=0.01$ and $K_2=0.03$, $\mu_Y$, $\mu_X$, $\sigma_Y^2$, $\sigma_X^2$ and $\sigma_{YX}$ are the means of $Y$ and $X$, variances of $Y$ and $X$, and the covariance between $Y$ and $X$ respectively. Since the maximum value of SSIM is 1, the structural loss used to optimize DEAR-3D is expressed as follows:
\begin{equation}
L_{sl}=\frac{1}{N_b}\sum_{i=1}^{N_b}\Big [ 1-SSIM(Y_i,X_i)\Big ]. \label{eqn:7}
\end{equation}

\subsubsection{Adversarial Loss}
The adversarial used in DEAR-3D is for the generator to produce realistic images that are indistinguishable by the discriminator network. Refer to Eq. \eqref{eq3}, the adversarial loss is expressed as follows:
\begin{equation}
L_{al}=-\mathbb{E}_{S_{SV}}[D(G(S_{SV}))] \label{eqn:8}
\end{equation}
The overall objective function of $G$ is then expressed as follows:
\begin{equation}
L_G=\lambda_{al}\cdot L_{al}+\lambda_{sl}\cdot L_{sl}+L_2 \label{eqn:9}
\end{equation}
where $\lambda_{al}$ and $\lambda_{sl}$ are hyper-parameters to balance different loss functions.

\subsection{Corresponding 2D networks for comparisons}
To evaluate the performance of the proposed 3D network, a 2D network is built for bench-marking, which is denoted as DEAR-2D. DEAR-2D uses the exactly same structure as the DEAR-3D, except that all the 3D convolutional layers in the dense blocks are replaced with 2D convolutional layers. Please note that the number of parameters of DEAR-2D will be less than that of DEAR-3D due to the fact that the dimension of input 2D batches is significant smaller than the dimension of 3D batches. For a fair comparison, another 2D network is built with an accordingly increased number of training parameters, denoted as DEAR-2D-i. The number of training parameters is increased by increasing number of filters in 2D convolutional layers.  Different from DEAR-3D, the 2D counterparts only utilize 2D convolutional layers to extract 2D feature maps from a batch of 2D input images. Therefore, the 2D counterparts aims at reconstructing 2D images instead of 3D images, which may lead to a potential loss in contextual information. Consequently, in DEAR-2D, all the 2D convolutional layers in both the encoder-decoder part and the dense blocks contain 38 filters with kernel dimension $3\times 3$. On the other hand, in DEAR-2D-i, all the 2D convolutional layers in both the encoder-decoder part and the dense blocks contain 48 filters with kernel dimension $3\times 3$. Table \ref{tab1} shows the numbers of parameters of the three networks.

\begin{table}[htbp]
\centering
\caption{Number of parameters used in different networks}
\begin{tabular}{cccc}
\toprule
\textbf{\# Parameters}	& \textbf{DEAR-3D}	& \textbf{DEAR-2D} & \textbf{DEAR-2D-i}\\
\midrule
		& 5,123,617			& 3,459,749 & 5,519,329\\
\bottomrule
\end{tabular}
\label{tab1}
\end{table}

Moreover, to demonstrate the effectiveness of different loss functions used to optimize the proposed neural network, 2D and 3D networks with different combinations of loss components are considered for comparison.

\section{Experimental design and results}\label{sec:exp}
\subsection{Dataset and Pre-processing}

A clinical abdominal dataset was used to train and evaluate the performance of the proposed DEAR-3D method. The dataset was prepared and authorized by Mayo Clinic for ``\emph{the 2016 NIH-AAPM-Mayo Clinic\ Low Dose CT Grand Challenge}'' \cite{noauthor_low_nodate}. The dataset contains a total of 5,936 abdominal CT images selected with 1 mm slice thickness. All the images were reconstructed from 2,304 projections under 100 peak kilovoltage (kVp), which were used as the ground-truth images to train the proposed method. The distance between the x-ray source and the detector array is 1085.6 milimeters, and the distance between the x-ray source and the iso-center is 595 milimeters. The pixel size is 0.664 millimeters. All the images are of $512\times 512$. For data-preprocessing, pixel values of patient images were normalized to be between 0 and 1. During the training process, 4 patients (a total of 2,566 images) were used for training, and 6 patients (a total of 3370 images) for validation and testing. Patches with dimension $64\times 64$ were cropped with stride 32 from the whole images for data augmentation, resulting in a total of 502,936 2D training patches. 2D patches were used to train the DEAR-2D and DEAR-2D-i networks. 3D patches were extracted from the pre-processed 2D patches to train the DEAR-3D network. 3D patches were extracted with stride of 1 in the $N_s$ dimension. Then, the optimized networks are applicable to images with any image dimension since the proposed DEAR-3D network contains only convolutional layers. The fan-beam Radon transform and fan-beam inverse Radon transform were used to simulate 75-view few-view images. 75-view sinograms were synthesized from angles equally distributed over a full scan range.

\subsection{Hyperparameter Selection and Network Comparison}
In the experiments, all codes were implemented in the TensorFlow framework \cite{abadi_tensorflow:_2016} on an NVIDIA Titan RTX GPU. The Adam optimization method was implemented to optimize the training parameters \cite{kingma_adam:_2015} with $\beta_1 = 0.9$ and $\beta_2 = 0.999$. During the training process, a mini-batch size of 10 was selected, resulting the input with dimensionality of $10\times 9\times 64\times 64\times 1$. The hyperparameter $\lambda$ used to balance the Wasserstein distance and the gradient penalty was set as 10, as suggested in \cite{gulrajani_improved_2017-1}. The learning rate was initialized as $1\times 10^{-4}$, and decreased by a factor of 2 after each epoch. The hyperparameters $\lambda_{al}$ and $\lambda_{sl}$ were adjusted using the following steps. First, the proposed network was optimized using only the MSE loss. The testing results were treated as the baseline for fine-tuning the other 2 hyper-parameters. Then, the SSIM loss was added as part of the objective function. Finally, the adversarial loss was added, and the hyperparameter $\lambda_{al}$ was fine-tuned. Through this process, $\lambda_{sl}$ and $\lambda_{al}$ were set to 0.5 and 0.0025 respectively. Please note that $\lambda_{sl}$ and $\lambda_{al}$ were fine-tuned for the best SSIM values in the validation set. 

For qualitative comparison, the proposed DEAR-3D network was compared with two deep-learning-based methods for few-view CT image reconstruction, including the FBPConvNet method (a classic U-net \cite{ronneberger_u-net:_2015} with conveying paths to solve the CT problem \cite{jin_deep_2017}) and a CNN-based residual network \cite{cong_deep-learning-based_2019} (denoted as residual-CNN in this paper). The network settings were made the same as the default settings described in the original papers. The analytical FBP method was used as a baseline for comparison.

Moreover, to highlight the effectiveness of the proposed objective functions used in the 3D architecture, as shown in Table \ref{tab2}, 5 different networks with different combinations of objective functions were trained for comparison: (1) The DEAR-2D network with only MSE loss and without WGAN (denoted as DEAR-2D$_1$); (2) DEAR-2D with MSE and SSIM but without WGAN (denoted as DEAR-2D$_2$); (3) DEAR-2D-i with MSE and SSIM loss and without WGAN (denoted as DEAR-2D-i); (4) DEAR-3D with MSE and SSIM loss but without WGAN (denoted as DEAR-3D$_1$); (5) a full DEAR-3D network with WGAN (denoted as DEAR-3D). Hyparameters for all of these 5 networks were experimentally adjusted using the steps mentioned above.

\begin{table}[htbp]
\centering
\caption{Summary of deep learning-based network architecture and
their optimization objective functions for few-view de-artifact
methods. The abbreviations for objection functions are MSE, SSIM and AL in this table are for
mean squared error, structural similarity index and adversarial loss respectively}
\begin{tabular}{cccc}
\midrule  & \textbf{MSE} & \textbf{SSIM} & \textbf{AL} \\
\midrule  DEAR-2D$_1$ & \checkmark & & \\
\midrule  DEAR-2D$_2$ & \checkmark & \checkmark & \\
\midrule  DEAR-2D-i & \checkmark & \checkmark & \\
\midrule DEAR-3D$_1$ & \checkmark & \checkmark & \\
\midrule DEAR-3D & \checkmark & \checkmark & \checkmark\\
\midrule
\end{tabular}
\label{tab2}
\end{table}

\subsection{Comparison with Other Deep-learning methods}

\begin{figure*}[htbp]
\centering
\includegraphics[width=\textwidth]{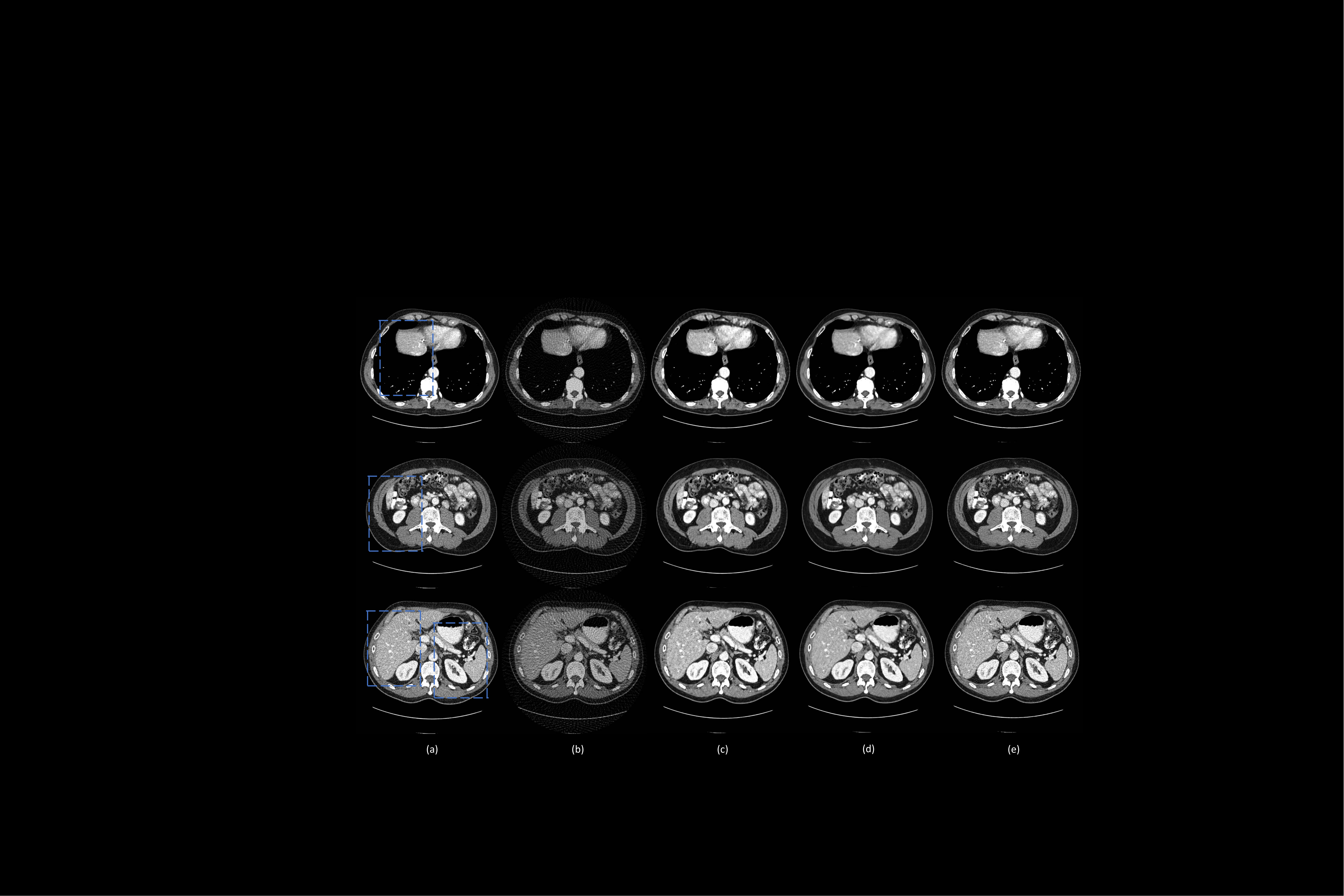}
\caption{Representative images reconstructed using different methods. (a) The ground-truth, (b) FBP, (c) residual-CNN, (d)FBPConvNet, and (e)DEAR-3D methods. The blue boxes mark the Region of Interests (ROIs). The display window is set as [-160, 240] HU for better visualizing lesions and subtle details.}
\label{fig:Mayo_testing}
\end{figure*}

\begin{figure}[htbp]
\centering
\includegraphics[width=0.6\textwidth]{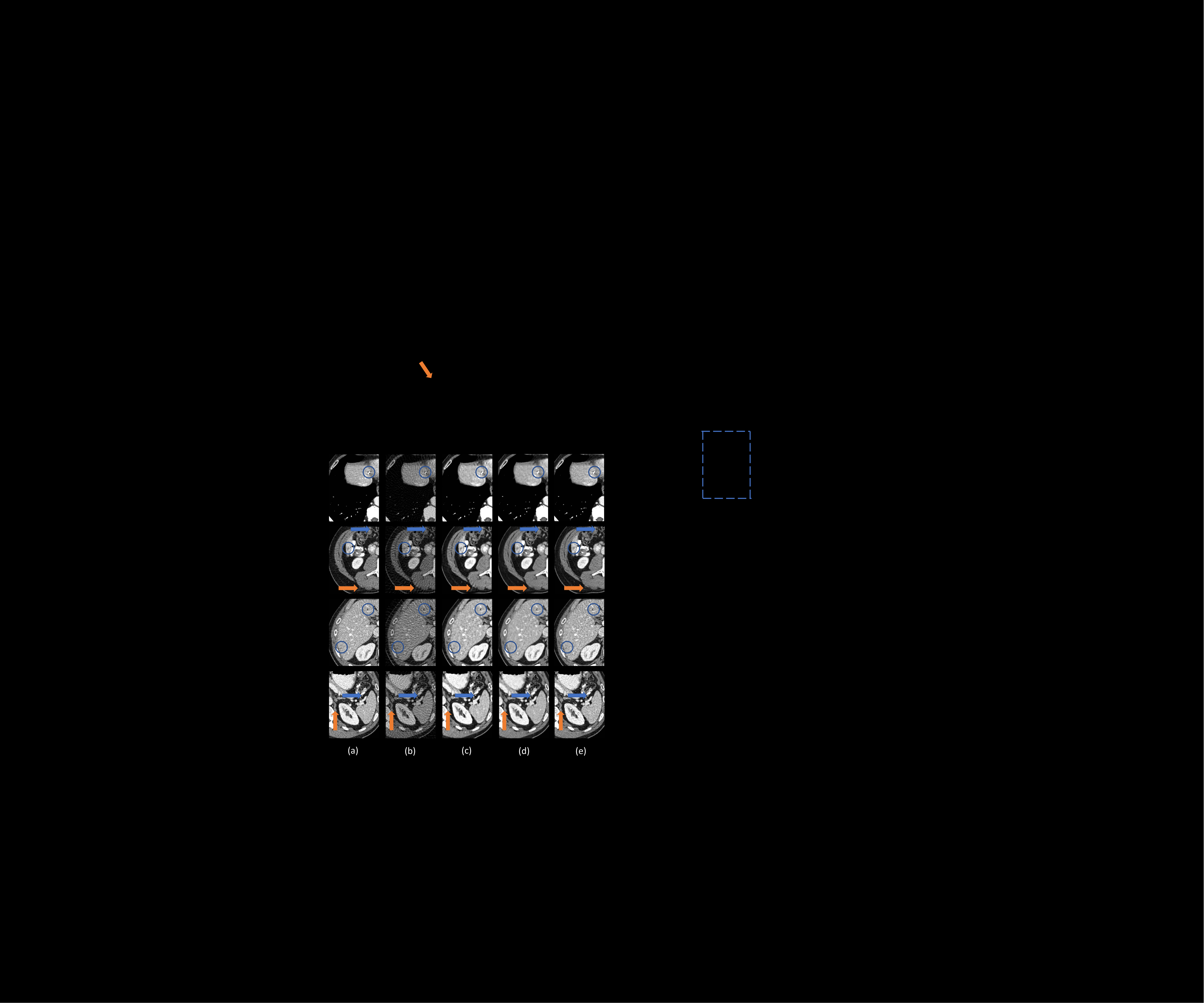}
\caption{Zoomed-in ROIs (The blue boxes in Fig. \ref{fig:Mayo_testing}). (a) The ground-truth, (b) FBP, (c) residual-CNN, (d)FBPConvNet, and (e)DEAR-3D. The blue and orange circles mark lesion locations. The blue and orange arrows indicate some subtle details. The display window is [-300, 300] HU.}
\label{fig:ROI}
\end{figure}

To visualize the performance of different methods, a few representative slices were selected from the testing set. Fig. \ref{fig:Mayo_testing} shows results using different methods from 75-view few-view images. Three metrics, peak signal-to-noise ratio (PSNR) \cite{korhonen_peak_2012}, SSIM, and root-mean-square-error (RMSE)\cite{willmott_advantages_2005} were computed for quantitative assessment. The quantitative results are shown in Table \ref{tab3}. For better evaluation of the image quality, the regions-of-interest (ROIs) are marked by rectangles in Fig. \ref{fig:Mayo_testing} are magnified in Fig. \ref{fig:ROI}.

\begin{table*}[htbp]
\centering
\caption{Quantitative measurements on different methods ($MEAN \pm STD$). For each metric, the best result is mark as bold. The measurements were obtained by averaging the values in the testing set.}
\setlength{\textwidth}{3pt}
\begin{tabular}{ccccc}
\midrule  & \textbf{FBP} & \textbf{FBPConvNet} & \textbf{residual-CNN} & \textbf{DEAR-3D} \\
\midrule
PSNR & $ 25.238\pm 0.967$ & $31.437\pm 1.366$ & $30.412\pm 1.269$ & \boldmath{$32.418 \pm 1.393$}  \\
\midrule
SSIM & $0.550 \pm 0.031$ & $0.871\pm 0.034$ & $0.870\pm 0.035 $ & \boldmath{$0.878 \pm 0.033$}  \\
\midrule
RMSE &  $0.055\pm 0.006 $ & $0.027 \pm 0.004$ & $0.030\pm 0.004$ & \boldmath{$0.025\pm 0.005$}  \\
\midrule
\end{tabular}
\label{tab3}
\end{table*}
\begin{figure*}[htbp]
    \centering
    \includegraphics[width=\textwidth]{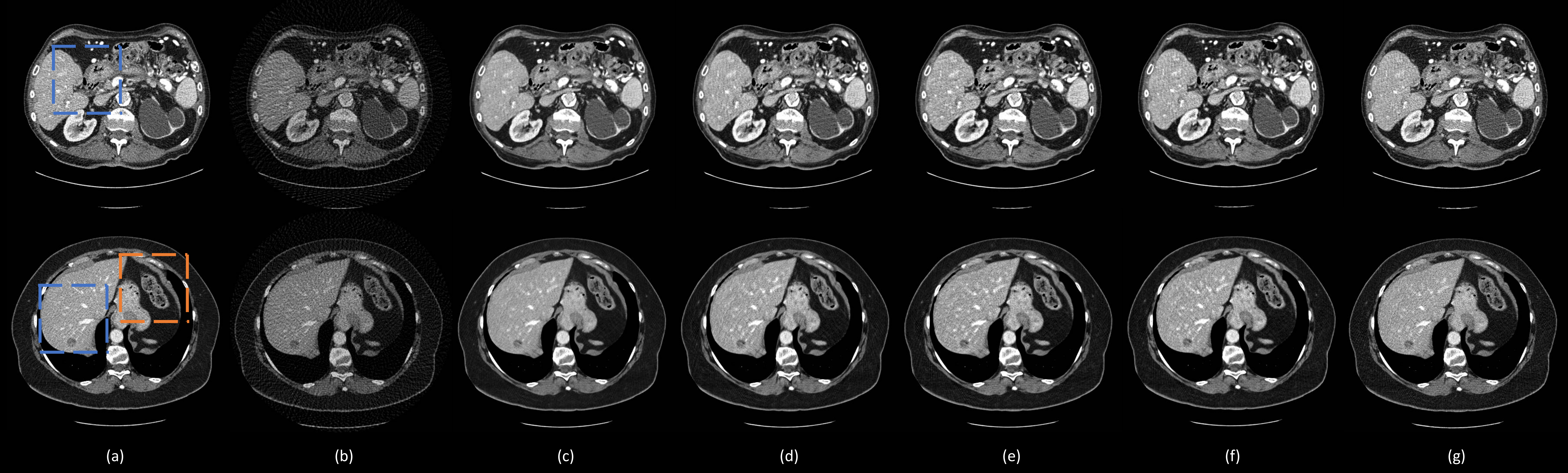}
    \caption{Representative images reconstructed using different methods. (a) The ground-truth, (b)FBP, (c) DEAR2D$_1$, (d) DEAR-2D$_2$, (e)DEAR-2D-i, (f) DEAR-3D$_1$, and (g) DEAE-3D. The blue boxes mark the region of interests (ROIs). The orange boxes contain subtle details in the images. The display window is set as [-160, 240] HU for better visualizing lesions and subtle details.}
    \label{fig:Mayo_testing2}
\end{figure*}

\begin{figure*}[htbp]
    \centering
    \includegraphics[width=\textwidth]{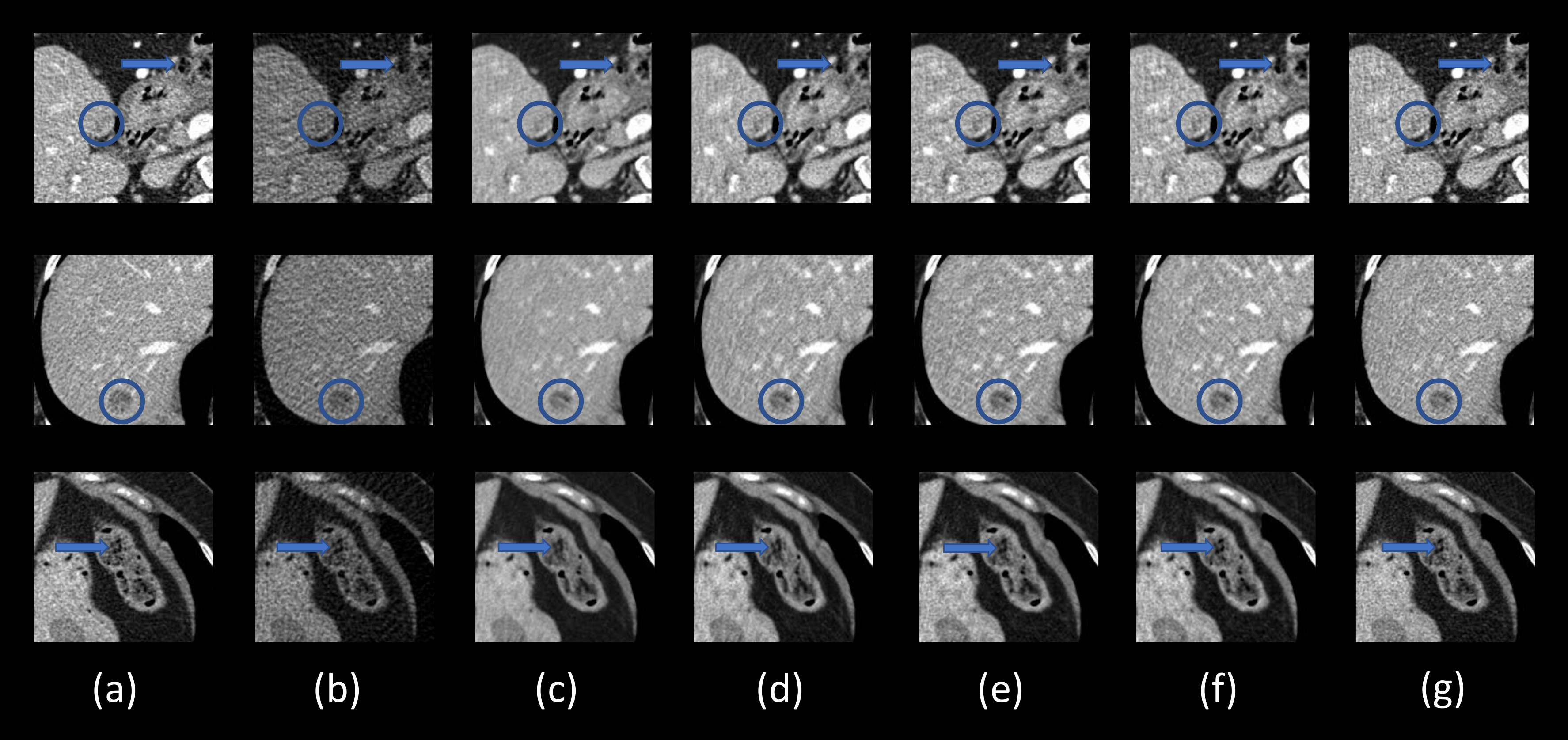}
    \caption{Zoomed-in areas of the lesion (The blue and orange boxes in Fig. \ref{fig:Mayo_testing2}). (a) The ground-truth, (b) FBP, (c) DEAR-2D$_1$, (d) DEAR-2D$_2$, (e) DEAR-2D-i, (f) DEAR-3D$_1$, and (g) DEAE-3D. The blue circles mark lesion locations. The blue arrows indicate some subtle details. The display window is set as [-160, 240] HU for better visualizing lesions and subtle details.}
    \label{fig:ROI2}
\end{figure*}

\begin{table*}[htbp]
\centering
\caption{Quantitative measurements for different methods ($MEAN \pm STD$). For each metric, the best result is mark as bold. Measurements were obtained by averaging the values in the testing dataset. (b) FBP, (c) DEAR-2D$_1$, (d) DEAR-2D$_2$, (e) DEAR-2D-i, (f) DEAR-3D$_1$, and (g) DEAR-3D.}
\setlength{\textwidth}{3pt}
\begin{tabular}{ccccccc}
\midrule & \textbf{FBP} &  \textbf{DEAR-2D$_1$} & \textbf{DEAR-2D$_2$} & \textbf{DEAR-2D-i} & \textbf{DEAR-3D$_1$} & \textbf{DEAR-3D}\\
\midrule
PSNR & $25.238\pm 0.967$ & $31.556\pm 1.378$ & $31.043\pm 1.360$ & $30.938\pm 1.381$ & $31.068 \pm 1.355$ & \boldmath{$32.418 \pm 1.393$}  \\
\midrule
SSIM & $0.550\pm 0.031$ & $0.861 \pm 0.033$ & $0.868\pm 0.034$ & $0.865\pm 0.033 $ & $0.873 \pm 0.033$ & \boldmath{$0.878 \pm 0.033$}\\
\midrule
RMSE & $0.055\pm 0.006$ & $0.027\pm 0.004 $ & $0.028 \pm 0.004$ & $0.029\pm 0.005$ & $0.027\pm 0.004$ & \boldmath{$0.025\pm 0.005$} \\
\midrule
\end{tabular}
\label{tab4}
\end{table*}

The ground-truth images and the corresponding few-view images are presented in Fig. \ref{fig:Mayo_testing}a and \ref{fig:Mayo_testing}b respectively. As shown in Fig. \ref{fig:Mayo_testing}b, streak artifacts are clearly visible in the images reconstructed using the FBP method. As shown in the ground-truth images in Fig.\ref{fig:Mayo_testing}a, lesions and subtle details are visible which are hidden by few-view artifacts in Fig. \ref{fig:Mayo_testing}a. The results from the 2D-based deep-learning reconstruction methods (FBPConvNet and residual-CNN) are shown in Fig. \ref{fig:Mayo_testing}c and \ref{fig:Mayo_testing}d as well as Fig. \ref{fig:ROI}c and \ref{fig:ROI}d respectively. These 2D methods can effectively reduce artifacts but they would potentially miss spatial correlation between adjacent slices, resulting in loss of subtle but critical details. As shown in the first and second row of Fig. \ref{fig:ROI}, FBPConvNet and residual-CNN tend to distort or smooth out some subtle details in the ROIs but these details are visible in the full-dose images reconstructed by the FBP method (indicated by the blue and orange arrows in Fig. \ref{fig:ROI}). Moreover, it is observed that, residual-CNN is unable to effectively remove streak artifacts in the reconstructed images, especially along the boundaries (the orange and blue arrows in Fig. \ref{fig:ROI}). Our proposed method, DEAR-3D, is better at removing artifacts as well as keeping tiny but vital details than the competitive methods. The proposed DEAR-3D method is also better at recovering image texture than the other methods, this may be due to the processing capability of the 3D network and the discriminative power of the WGAN framework.

\subsection{Ablation Analysis}

This subsection demonstrate the effectiveness of different components in the proposed DEAR-3D network. As mentioned above, 5 variants of the DEAR-3D network were trained for this purpose. The results are shown in Fig. \ref{fig:Mayo_testing2}, with the corresponding quantitative measurements in Table \ref{tab4}. The zoomed-in regions-of-interest (ROIs), which are marked by rectangles in Fig. \ref{fig:Mayo_testing2}, are shown in Fig. \ref{fig:ROI2}. As presented in Fig. \ref{fig:ROI2}, due to the improper 2D design of the objective function, the DEAR-2D network with only MSE loss tends to smooth out features such as the lesion, leading to an unacceptable image quality (the lesion becomes barely visible in the first row in Fig. \ref{fig:ROI2}). Adding SSIM as part of the objective function improved the overall image quality but due to the lack of 3D spatial context, the 2D based methods are unable to recover subtle details (indicated by the blue arrows in the first and third rows in Fig. \ref{fig:ROI2}). There is no significant difference observed between the DEAR-2D and the DEAR-2D-i networks. Lastly, the combination of the 3D architecture, WGAN and the adversarial loss improved image texture and overall image quality, which is desirable in practice. In summary, it is observed that the 2D-based methods compromise some details in the reconstructed images (the blue arrows in Fig. \ref{fig:ROI2}), and by providing information from adjacent slices, the DEAR-3D network performs better than the other methods at removing artifacts and keeping image texture.

\section{Discussion and Conclusions}\label{sec:conclusion}

Few-view CT may be implemented as a mechanically stationary scanner in the future \cite{cramer_stationary_2018} for health-care and other utilities. Current commercial CT scanners use one or two x-ray sources mounted on a rotating gantry, and take hundreds of projections around a patient. The rotating mechanism is not only massive but also power-consuming. Hence, current commercial CT scanners are inaccessible outside hospitals and imaging centers, due to their size, weight, and cost. Designing a stationary gantry with multiple miniature x-ray sources is an interesting approach to resolve this issue \cite{cramer_stationary_2018}. Unfortunately, the current technology does not allow us to assemble hundreds of miniature x-ray sources in a ring for reconstructing a high-quality CT image over an ROI of a decent aperture. Few-view CT is an attractive option. However, streak artifacts would be introduced from a few-view scan due to insufficiency of projection data. Recently, deep learning has achieved remarkable results for few-view CT, and our proposed DEAR-3D network is a step forward along this direction.

This paper has introduced a novel 3D Deep Encoder-decoder Adversarial Reconstruction Network (DEAR-3D) for directly reconstructing a 3D volume from 3D input data. Compared with 2D-based methods \cite{jin_deep_2017, lee_deep-neural-network-based_2019, chen_learn:_2018, xie_dual_2019-1, cong_deep-learning-based_2019}, DEAR-3D avoids the potential loss in the 3D spatial context. Specifically, our proposed network is featured by (1) a 3D convolutional encoder-decoder network with conveying-paths; (2) the Wasserstein GAN framework for optimal parameters; and (3) the powerful DenseNet architecture for improved performance.

In conclusion, we have presented a novel 3D deep network, DEAR-3D,  for solving the few-view CT problem. The proposed method outperforms 3D deep-learning methods and promises clinical utilities such as  breast cone-beam CT and C-arm cone-beam CT for future research probabilities. In the follow-up investigation, we plan to further improve the network and  perform more experiments to optimize and validate the DEAR-3D network.



\authorcontributions{H.Xie. and H.Shan. initiated the project and designed the experiments. H.Xie. performed machine learning research and experiments. H.Xie. wrote the paper. H.Shan. and G.Wang.
participated in the discussions and edited the paper.}

\reftitle{References}

\externalbibliography{yes}
\bibliography{main}

\begin{thebibliography}{-------}
\providecommand{\natexlab}[1]{#1}

\bibitem[Brenner and Hall(2007)]{brenner_computed_2007}
Brenner, D.J.; Hall, E.J.
\newblock Computed {Tomography} — {An} {Increasing} {Source} of {Radiation}
  {Exposure}.
\newblock {\em N Engl J Med} {\bf 2007}, {\em 357},~2277--2284.
\newblock
  doi:{\changeurlcolor{black}\href{https://doi.org/10.1056/NEJMra072149}{\detokenize{10.1056/NEJMra072149}}}.

\bibitem[Shan \em{et~al.}(2018)Shan, Zhang, Yang, Kruger, Kalra, Sun, Cong, and
  Wang]{shan_3-d_2018-1}
Shan, H.; Zhang, Y.; Yang, Q.; Kruger, U.; Kalra, M.K.; Sun, L.; Cong, W.;
  Wang, G.
\newblock 3-{D} {Convolutional} {Encoder}-{Decoder} {Network} for {Low}-{Dose}
  {CT} via {Transfer} {Learning} {From} a 2-{D} {Trained} {Network}.
\newblock {\em IEEE Trans Med Imaging} {\bf 2018}, {\em 37},~1522--1534.
\newblock
  doi:{\changeurlcolor{black}\href{https://doi.org/10.1109/TMI.2018.2832217}{\detokenize{10.1109/TMI.2018.2832217}}}.

\bibitem[Chen \em{et~al.}(2017)Chen, Zhang, Zhang, Liao, Li, Zhou, and
  Wang]{chen_low-dose_2017-1}
Chen, H.; Zhang, Y.; Zhang, W.; Liao, P.; Li, K.; Zhou, J.; Wang, G.
\newblock Low-dose {CT} via convolutional neural network.
\newblock {\em Biomed Opt Express} {\bf 2017}, {\em 8},~679--694.
\newblock
  doi:{\changeurlcolor{black}\href{https://doi.org/10.1364/BOE.8.000679}{\detokenize{10.1364/BOE.8.000679}}}.

\bibitem[Shan \em{et~al.}(2019)Shan, Padole, Homayounieh, Kruger, Khera,
  Nitiwarangkul, Kalra, and Wang]{shan_competitive_2019-1}
Shan, H.; Padole, A.; Homayounieh, F.; Kruger, U.; Khera, R.D.; Nitiwarangkul,
  C.; Kalra, M.K.; Wang, G.
\newblock Competitive performance of a modularized deep neural network compared
  to commercial algorithms for low-dose {CT} image reconstruction.
\newblock {\em Nat. Mach. Intell} {\bf 2019}, {\em 1},~269--276.
\newblock
  doi:{\changeurlcolor{black}\href{https://doi.org/10.1038/s42256-019-0057-9}{\detokenize{10.1038/s42256-019-0057-9}}}.

\bibitem[Chen \em{et~al.}(2017)Chen, Zhang, Kalra, Lin, Chen, Liao, Zhou, and
  Wang]{chen_low-dose_2017}
Chen, H.; Zhang, Y.; Kalra, M.K.; Lin, F.; Chen, Y.; Liao, P.; Zhou, J.; Wang,
  G.
\newblock Low-{Dose} {CT} with a {Residual} {Encoder}-{Decoder} {Convolutional}
  {Neural} {Network} ({RED}-{CNN}).
\newblock {\em IEEE Trans Med Imaging} {\bf 2017}, {\em 36},~2524--2535.
\newblock
  doi:{\changeurlcolor{black}\href{https://doi.org/10.1109/TMI.2017.2715284}{\detokenize{10.1109/TMI.2017.2715284}}}.

\bibitem[Glick(2007)]{glick_breast_2007-1}
Glick, S.J.
\newblock Breast {CT}.
\newblock {\em Annu. Rev. Biomed. Eng.} {\bf 2007}, {\em 9},~501--526.
\newblock
  doi:{\changeurlcolor{black}\href{https://doi.org/10.1146/annurev.bioeng.9.060906.151924}{\detokenize{10.1146/annurev.bioeng.9.060906.151924}}}.

\bibitem[Wallace \em{et~al.}(2008)Wallace, Kuo, Glaiberman, Binkert, Orth, and
  Soulez]{wallace_three-dimensional_2008}
Wallace, M.J.; Kuo, M.D.; Glaiberman, C.; Binkert, C.A.; Orth, R.C.; Soulez, G.
\newblock Three-{Dimensional} {C}-arm {Cone}-beam {CT}: {Applications} in the
  {Interventional} {Suite}.
\newblock {\em J Vasc Interv Radiol} {\bf 2008}, {\em 19},~799--813.
\newblock
  doi:{\changeurlcolor{black}\href{https://doi.org/10.1016/j.jvir.2008.02.018}{\detokenize{10.1016/j.jvir.2008.02.018}}}.

\bibitem[Jerri(1977)]{jerri_shannon_1977}
Jerri, A.J.
\newblock The {Shannon} sampling theorem—{Its} various extensions and
  applications: {A} tutorial review.
\newblock {\em Proceedings of the IEEE} {\bf 1977}, {\em 65},~1565--1596.
\newblock
  doi:{\changeurlcolor{black}\href{https://doi.org/10.1109/PROC.1977.10771}{\detokenize{10.1109/PROC.1977.10771}}}.

\bibitem[Wang \em{et~al.}(2007)Wang, Ye, and Yu]{wang_approximate_2007}
Wang, G.; Ye, Y.; Yu, H.
\newblock Approximate and exact cone-beam reconstruction with standard and
  non-standard spiral scanning.
\newblock {\em Phys Med Biol} {\bf 2007}, {\em 52},~R1--13.
\newblock
  doi:{\changeurlcolor{black}\href{https://doi.org/10.1088/0031-9155/52/6/R01}{\detokenize{10.1088/0031-9155/52/6/R01}}}.

\bibitem[Gordon \em{et~al.}(1970)Gordon, Bender, and
  Herman]{gordon_algebraic_1970}
Gordon, R.; Bender, R.; Herman, G.T.
\newblock Algebraic reconstruction techniques ({ART}) for three-dimensional
  electron microscopy and x-ray photography.
\newblock {\em Journal of theoretical biology} {\bf 1970}, {\em 29},~471--481.

\bibitem[Andersen(1984)]{andersen_simultaneous_1984}
Andersen, A.
\newblock Simultaneous {Algebraic} {Reconstruction} {Technique} ({SART}): {A}
  superior implementation of the {ART} algorithm.
\newblock {\em Ultrasonic Imaging} {\bf 1984}, {\em 6},~81--94.
\newblock
  doi:{\changeurlcolor{black}\href{https://doi.org/10.1016/0161-7346(84)90008-7}{\detokenize{10.1016/0161-7346(84)90008-7}}}.

\bibitem[Dempster \em{et~al.}(1977)Dempster, Laird, and
  work(s):]{dempster_maximum_1977}
Dempster, A.P.; Laird, N.M.; work(s):, D.B.R.R.
\newblock Maximum {Likelihood} from {Incomplete} {Data} via the {EM}
  {Algorithm}.
\newblock {\em Journal of the Royal Statistical Society. Series B
  (Methodological)} {\bf 1977}, {\em 39},~1--38.

\bibitem[Wang(2016)]{wang_perspective_2016}
Wang, G.
\newblock A {Perspective} on {Deep} {Imaging}.
\newblock {\em IEEE Access} {\bf 2016}, {\em 4},~8914--8924.
\newblock
  doi:{\changeurlcolor{black}\href{https://doi.org/10.1109/ACCESS.2016.2624938}{\detokenize{10.1109/ACCESS.2016.2624938}}}.

\bibitem[Wang \em{et~al.}(2015)Wang, Butler, Yu, and Campbell]{wang_guest_2015}
Wang, G.; Butler, A.; Yu, H.; Campbell, M.
\newblock Guest {Editorial} {Special} {Issue} on {Spectral} {CT}.
\newblock {\em IEEE Trans Med Imaging} {\bf 2015}, {\em 34},~693--696.
\newblock
  doi:{\changeurlcolor{black}\href{https://doi.org/10.1109/TMI.2015.2404591}{\detokenize{10.1109/TMI.2015.2404591}}}.

\bibitem[Greenspan \em{et~al.}(2016)Greenspan, Ginneken, and
  Summers]{greenspan_guest_2016}
Greenspan, H.; Ginneken, B.v.; Summers, R.M.
\newblock Guest {Editorial} {Deep} {Learning} in {Medical} {Imaging}:
  {Overview} and {Future} {Promise} of an {Exciting} {New} {Technique}.
\newblock {\em IEEE Trans Med Imaging} {\bf 2016}, {\em 35},~1153--1159.
\newblock
  doi:{\changeurlcolor{black}\href{https://doi.org/10.1109/TMI.2016.2553401}{\detokenize{10.1109/TMI.2016.2553401}}}.

\bibitem[Jin \em{et~al.}(2017)Jin, McCann, Froustey, and Unser]{jin_deep_2017}
Jin, K.H.; McCann, M.T.; Froustey, E.; Unser, M.
\newblock Deep {Convolutional} {Neural} {Network} for {Inverse} {Problems} in
  {Imaging}.
\newblock {\em IEEE Trans. Image Process.} {\bf 2017}, {\em 26},~4509--4522.
\newblock
  doi:{\changeurlcolor{black}\href{https://doi.org/10.1109/TIP.2017.2713099}{\detokenize{10.1109/TIP.2017.2713099}}}.

\bibitem[Ronneberger \em{et~al.}(2015)Ronneberger, Fischer, and
  Brox]{ronneberger_u-net:_2015}
Ronneberger, O.; Fischer, P.; Brox, T.
\newblock U-{Net}: {Convolutional} {Networks} for {Biomedical} {Image}
  {Segmentation}.
\newblock {\em arXiv:1505.04597 [cs]} {\bf 2015}.
\newblock arXiv: 1505.04597.

\bibitem[Lee \em{et~al.}(2019)Lee, Lee, Kim, Cho, and
  Cho]{lee_deep-neural-network-based_2019}
Lee, H.; Lee, J.; Kim, H.; Cho, B.; Cho, S.
\newblock Deep-{Neural}-{Network}-{Based} {Sinogram} {Synthesis} for
  {Sparse}-{View} {CT} {Image} {Reconstruction}.
\newblock {\em IEEE Transactions on Radiation and Plasma Medical Sciences} {\bf
  2019}, {\em 3},~109--119.
\newblock
  doi:{\changeurlcolor{black}\href{https://doi.org/10.1109/TRPMS.2018.2867611}{\detokenize{10.1109/TRPMS.2018.2867611}}}.

\bibitem[Chen \em{et~al.}(2018)Chen, Zhang, Chen, Zhang, Zhang, Sun, Lv, Liao,
  Zhou, and Wang]{chen_learn:_2018}
Chen, H.; Zhang, Y.; Chen, Y.; Zhang, J.; Zhang, W.; Sun, H.; Lv, Y.; Liao, P.;
  Zhou, J.; Wang, G.
\newblock {LEARN}: {Learned} {Experts}’ {Assessment}-{Based} {Reconstruction}
  {Network} for {Sparse}-{Data} {CT}.
\newblock {\em IEEE Trans Med Imaging} {\bf 2018}, {\em 37},~1333--1347.
\newblock
  doi:{\changeurlcolor{black}\href{https://doi.org/10.1109/TMI.2018.2805692}{\detokenize{10.1109/TMI.2018.2805692}}}.

\bibitem[Lecun and Bengio(1995)]{lecun_convolutional_1995}
Lecun, Y.; Bengio, Y.
\newblock Convolutional networks for images, speech, and time-series.
\newblock {\em The handbook of brain theory and neural networks} {\bf 1995}.

\bibitem[Li \em{et~al.}(2019)Li, Li, Zhang, Montoya, and
  Chen]{li_learning_2019}
Li, Y.; Li, K.; Zhang, C.; Montoya, J.; Chen, G.
\newblock Learning to {Reconstruct} {Computed} {Tomography} ({CT}) {Images}
  {Directly} from {Sinogram} {Data} under {A} {Variety} of {Data} {Acquisition}
  {Conditions}.
\newblock {\em IEEE Trans Med Imaging} {\bf 2019}, pp. 1--1.
\newblock
  doi:{\changeurlcolor{black}\href{https://doi.org/10.1109/TMI.2019.2910760}{\detokenize{10.1109/TMI.2019.2910760}}}.

\bibitem[Xie \em{et~al.}(2019)Xie, Shan, Cong, Zhang, Liu, Ning, and
  Wang]{xie_dual_2019-1}
Xie, H.; Shan, H.; Cong, W.; Zhang, X.; Liu, S.; Ning, R.; Wang, G.
\newblock Dual network architecture for few-view {CT} - trained on {ImageNet}
  data and transferred for medical imaging.
\newblock  Developments in {X}-{Ray} {Tomography} {XII}. International Society
  for Optics and Photonics,  2019, Vol. 11113, p. 111130V.
\newblock
  doi:{\changeurlcolor{black}\href{https://doi.org/10.1117/12.2531198}{\detokenize{10.1117/12.2531198}}}.

\bibitem[Goodfellow \em{et~al.}(2014)Goodfellow, Pouget-Abadie, Mirza, Xu,
  Warde-Farley, Ozair, Courville, and Bengio]{goodfellow_generative_2014}
Goodfellow, I.J.; Pouget-Abadie, J.; Mirza, M.; Xu, B.; Warde-Farley, D.;
  Ozair, S.; Courville, A.; Bengio, Y.
\newblock Generative {Adversarial} {Networks}.
\newblock {\em arXiv:1406.2661 [cs, stat]} {\bf 2014}.
\newblock arXiv: 1406.2661.

\bibitem[Schaller \em{et~al.}(1998)Schaller, Flohr, and
  Steffen]{schaller_efficient_1998}
Schaller, S.; Flohr, T.; Steffen, P.
\newblock An efficient {Fourier} method for 3-{D} radon inversion in exact
  cone-beam {CT} reconstruction.
\newblock {\em IEEE Trans Med Imaging} {\bf 1998}, {\em 17},~244--250.
\newblock
  doi:{\changeurlcolor{black}\href{https://doi.org/10.1109/42.700736}{\detokenize{10.1109/42.700736}}}.

\bibitem[Grangeat(1991)]{grangeat_mathematical_1991}
Grangeat, P.
\newblock Mathematical framework of cone beam 3D reconstruction via the first
  derivative of the radon transform.
\newblock  Mathematical {Methods} in {Tomography}; Herman, G.T.; Louis, A.K.;
  Natterer, F., Eds. Springer Berlin Heidelberg,  1991, Lecture {Notes} in
  {Mathematics}, pp. 66--97.

\bibitem[Grangeat \em{et~al.}(1991)Grangeat, Masson, Melennec, and
  Sire]{grangeat_evaluation_1991}
Grangeat, P.P.; Masson, P.L.; Melennec, P.; Sire, P.
\newblock Evaluation of the 3-{D} radon transform algorithm for cone beam
  reconstruction.
\newblock  Medical {Imaging} {V}: {Image} {Processing}. International Society
  for Optics and Photonics,  1991, Vol. 1445, pp. 320--331.
\newblock
  doi:{\changeurlcolor{black}\href{https://doi.org/10.1117/12.45229}{\detokenize{10.1117/12.45229}}}.

\bibitem[Katsevich(2004)]{katsevich_improved_2004}
Katsevich, A.
\newblock An improved exact filtered backprojection algorithm for spiral
  computed tomography.
\newblock {\em Adv. Appl. Math.} {\bf 2004}, {\em 32},~681--697.
\newblock
  doi:{\changeurlcolor{black}\href{https://doi.org/10.1016/s0196-8858(03)00099-x}{\detokenize{10.1016/s0196-8858(03)00099-x}}}.

\bibitem[Arjovsky \em{et~al.}(2017)Arjovsky, Chintala, and
  Bottou]{arjovsky_wasserstein_2017}
Arjovsky, M.; Chintala, S.; Bottou, L.
\newblock Wasserstein {Generative} {Adversarial} {Networks}.
\newblock  International {Conference} on {Machine} {Learning},  2017, pp.
  214--223.

\bibitem[Gulrajani \em{et~al.}(2017)Gulrajani, Ahmed, Arjovsky, Dumoulin, and
  Courville]{gulrajani_improved_2017-1}
Gulrajani, I.; Ahmed, F.; Arjovsky, M.; Dumoulin, V.; Courville, A.C.
\newblock Improved {Training} of {Wasserstein} {GANs}. In {\em Advances in
  {Neural} {Information} {Processing} {Systems} 30}; Guyon, I.; Luxburg, U.V.;
  Bengio, S.; Wallach, H.; Fergus, R.; Vishwanathan, S.; Garnett, R., Eds.;
  Curran Associates, Inc.,  2017; pp. 5767--5777.

\bibitem[Donoho(2006)]{donoho_compressed_2006}
Donoho, D.L.
\newblock Compressed sensing.
\newblock {\em IEEE Transactions on Information Theory} {\bf 2006}, {\em
  52},~1289--1306.
\newblock
  doi:{\changeurlcolor{black}\href{https://doi.org/10.1109/TIT.2006.871582}{\detokenize{10.1109/TIT.2006.871582}}}.

\bibitem[Huang \em{et~al.}(2016)Huang, Liu, van~der Maaten, and
  Weinberger]{huang_densely_2016}
Huang, G.; Liu, Z.; van~der Maaten, L.; Weinberger, K.Q.
\newblock Densely {Connected} {Convolutional} {Networks}.
\newblock {\em arXiv:1608.06993 [cs]} {\bf 2016}.
\newblock arXiv: 1608.06993.

\bibitem[He \em{et~al.}(2015)He, Zhang, Ren, and Sun]{he_deep_2015}
He, K.; Zhang, X.; Ren, S.; Sun, J.
\newblock Deep {Residual} {Learning} for {Image} {Recognition}.
\newblock {\em arXiv:1512.03385 [cs]} {\bf 2015}.
\newblock arXiv: 1512.03385.

\bibitem[Wolterink \em{et~al.}(2017)Wolterink, Leiner, Viergever, and
  Išgum]{wolterink_generative_2017}
Wolterink, J.M.; Leiner, T.; Viergever, M.A.; Išgum, I.
\newblock Generative {Adversarial} {Networks} for {Noise} {Reduction} in
  {Low}-{Dose} {CT}.
\newblock {\em IEEE Trans Med Imaging} {\bf 2017}, {\em 36},~2536--2545.
\newblock
  doi:{\changeurlcolor{black}\href{https://doi.org/10.1109/TMI.2017.2708987}{\detokenize{10.1109/TMI.2017.2708987}}}.

\bibitem[Wang and Bovik(2009)]{wang_mean_2009}
Wang, Z.; Bovik, A.C.
\newblock Mean squared error: {Love} it or leave it? {A} new look at {Signal}
  {Fidelity} {Measures}.
\newblock {\em IEEE Signal Processing Magazine} {\bf 2009}, {\em 26},~98--117.
\newblock
  doi:{\changeurlcolor{black}\href{https://doi.org/10.1109/MSP.2008.930649}{\detokenize{10.1109/MSP.2008.930649}}}.

\bibitem[Zhao \em{et~al.}(2017)Zhao, Gallo, Frosio, and
  Kautz]{zhao_loss_2017-1}
Zhao, H.; Gallo, O.; Frosio, I.; Kautz, J.
\newblock Loss {Functions} for {Image} {Restoration} {With} {Neural}
  {Networks}.
\newblock {\em IEEE Trans. Comput. Imag} {\bf 2017}, {\em 3},~47--57.
\newblock
  doi:{\changeurlcolor{black}\href{https://doi.org/10.1109/TCI.2016.2644865}{\detokenize{10.1109/TCI.2016.2644865}}}.

\bibitem[{Zhou Wang} \em{et~al.}(2004){Zhou Wang}, Bovik, Sheikh, and
  Simoncelli]{zhou_wang_image_2004}
{Zhou Wang}.; Bovik, A.C.; Sheikh, H.R.; Simoncelli, E.P.
\newblock Image quality assessment: from error visibility to structural
  similarity.
\newblock {\em IEEE Trans. Image Process.} {\bf 2004}, {\em 13},~600--612.
\newblock
  doi:{\changeurlcolor{black}\href{https://doi.org/10.1109/TIP.2003.819861}{\detokenize{10.1109/TIP.2003.819861}}}.

\bibitem[noa()]{noauthor_low_nodate}
Low {Dose} {CT} {Grand} {Challenge}.
\newblock \url{https://www.aapm.org/grandchallenge/lowdosect/}.
\newblock Accessed: 2019-04-18.

\bibitem[Abadi \em{et~al.}(2016)Abadi, Agarwal, Barham, Brevdo, Chen, Citro,
  Corrado, Davis, Dean, Devin, Ghemawat, Goodfellow, Harp, Irving, Isard, Jia,
  Jozefowicz, Kaiser, Kudlur, Levenberg, Mane, Monga, Moore, Murray, Olah,
  Schuster, Shlens, Steiner, Sutskever, Talwar, Tucker, Vanhoucke, Vasudevan,
  Viegas, Vinyals, Warden, Wattenberg, Wicke, Yu, and
  Zheng]{abadi_tensorflow:_2016}
Abadi, M.; Agarwal, A.; Barham, P.; Brevdo, E.; Chen, Z.; Citro, C.; Corrado,
  G.S.; Davis, A.; Dean, J.; Devin, M.; Ghemawat, S.; Goodfellow, I.; Harp, A.;
  Irving, G.; Isard, M.; Jia, Y.; Jozefowicz, R.; Kaiser, L.; Kudlur, M.;
  Levenberg, J.; Mane, D.; Monga, R.; Moore, S.; Murray, D.; Olah, C.;
  Schuster, M.; Shlens, J.; Steiner, B.; Sutskever, I.; Talwar, K.; Tucker, P.;
  Vanhoucke, V.; Vasudevan, V.; Viegas, F.; Vinyals, O.; Warden, P.;
  Wattenberg, M.; Wicke, M.; Yu, Y.; Zheng, X.
\newblock {TensorFlow}: {Large}-{Scale} {Machine} {Learning} on {Heterogeneous}
  {Distributed} {Systems}.
\newblock {\em arXiv:1603.04467 [cs]} {\bf 2016}.
\newblock arXiv: 1603.04467.

\bibitem[Kingma and Ba(2015)]{kingma_adam:_2015}
Kingma, D.P.; Ba, L.J.
\newblock Adam: {A} {Method} for {Stochastic} {Optimization} {\bf 2015}.

\bibitem[Cong \em{et~al.}(2019)Cong, Shan, Zhang, Liu, Ning, and
  Wang]{cong_deep-learning-based_2019}
Cong, W.; Shan, H.; Zhang, X.; Liu, S.; Ning, R.; Wang, G.
\newblock Deep-learning-based breast {CT} for radiation dose reduction.
\newblock  Developments in {X}-{Ray} {Tomography} {XII}. International Society
  for Optics and Photonics,  2019, Vol. 11113, p. 111131L.
\newblock
  doi:{\changeurlcolor{black}\href{https://doi.org/10.1117/12.2530234}{\detokenize{10.1117/12.2530234}}}.

\bibitem[Korhonen and You(2012)]{korhonen_peak_2012}
Korhonen, J.; You, J.
\newblock Peak signal-to-noise ratio revisited: {Is} simple beautiful?
\newblock  2012 {Fourth} {International} {Workshop} on {Quality} of
  {Multimedia} {Experience},  2012, pp. 37--38.
\newblock
  doi:{\changeurlcolor{black}\href{https://doi.org/10.1109/QoMEX.2012.6263880}{\detokenize{10.1109/QoMEX.2012.6263880}}}.

\bibitem[Willmott and Matsuura(2005)]{willmott_advantages_2005}
Willmott, C.J.; Matsuura, K.
\newblock Advantages of the mean absolute error ({MAE}) over the root mean
  square error ({RMSE}) in assessing average model performance.
\newblock {\em Climate Research} {\bf 2005}, {\em 30},~79--82.
\newblock
  doi:{\changeurlcolor{black}\href{https://doi.org/10.3354/cr030079}{\detokenize{10.3354/cr030079}}}.

\bibitem[Cramer \em{et~al.}(2018)Cramer, Hecla, Wu, Lai, Boers, Yang, Moulton,
  Kenyon, Arzoumanian, Krull, Gendreau, and Gupta]{cramer_stationary_2018}
Cramer, A.; Hecla, J.; Wu, D.; Lai, X.; Boers, T.; Yang, K.; Moulton, T.;
  Kenyon, S.; Arzoumanian, Z.; Krull, W.; Gendreau, K.; Gupta, R.
\newblock Stationary {Computed} {Tomography} for {Space} and other
  {Resource}-constrained {Environments}.
\newblock {\em Sci Rep} {\bf 2018}, {\em 8}.
\newblock
  doi:{\changeurlcolor{black}\href{https://doi.org/10.1038/s41598-018-32505-z}{\detokenize{10.1038/s41598-018-32505-z}}}.

\end{thebibliography}



\end{document}